\documentstyle[11pt,newpasp,twoside,epsf]{article}

\makeatletter
\def\@jourvol{?}
\def\cpr@year{2002?}
\def\vol@title{A New Era in Cosmology}
\def\vol@author{N. Metcalfe, T. Shanks}
\makeatother

\newif\ifbw
\bwfalse

\newif\ifsmall
\smalltrue

\begin{document}

\title{{\sc\bf LensCLEAN}ing JVAS~B0218+357 to determine \boldmath$H_0$}
\author{Olaf Wucknitz}
\affil{Jodrell Bank Observatory, Manchester, UK \& Hamburger
  Sternwarte, Germany, {\tt ow@jb.man.ac.uk}}

\begin{abstract}
We use the radio ring in JVAS~B0218+357 to constrain the mass models
with LensCLEAN. $H_0$ is determined from the resulting model and the
time delay. B0218+357 is one of the best systems to reduce
systematic errors in $H_0$ and may thus be called a ``golden lens''.
\end{abstract}

\section{Introduction}

Gravitational lensing offers the unique possibility to determine the
Hubble constant $H_0$ (Refsdal 1964) involving the 
understanding of very little astrophysics, keeping statistical and
systematic errors well under control.
The main uncertainty is introduced by the mass models of the
lens. For multiply imaged compact sources, the number of constraints
is very small. To overcome this difficulty, lensed extended sources
can be used.

\section{The 18 karat golden lens JVAS~B0218+357}

\begin{center}\begin{minipage}{0.72\textwidth}
\ifbw%
\ifsmall%
\plotone{fig0218_bw2.eps}
\else%
\plotone{fig0218_bw.eps}
\fi%
\else%
\ifsmall%
\plotone{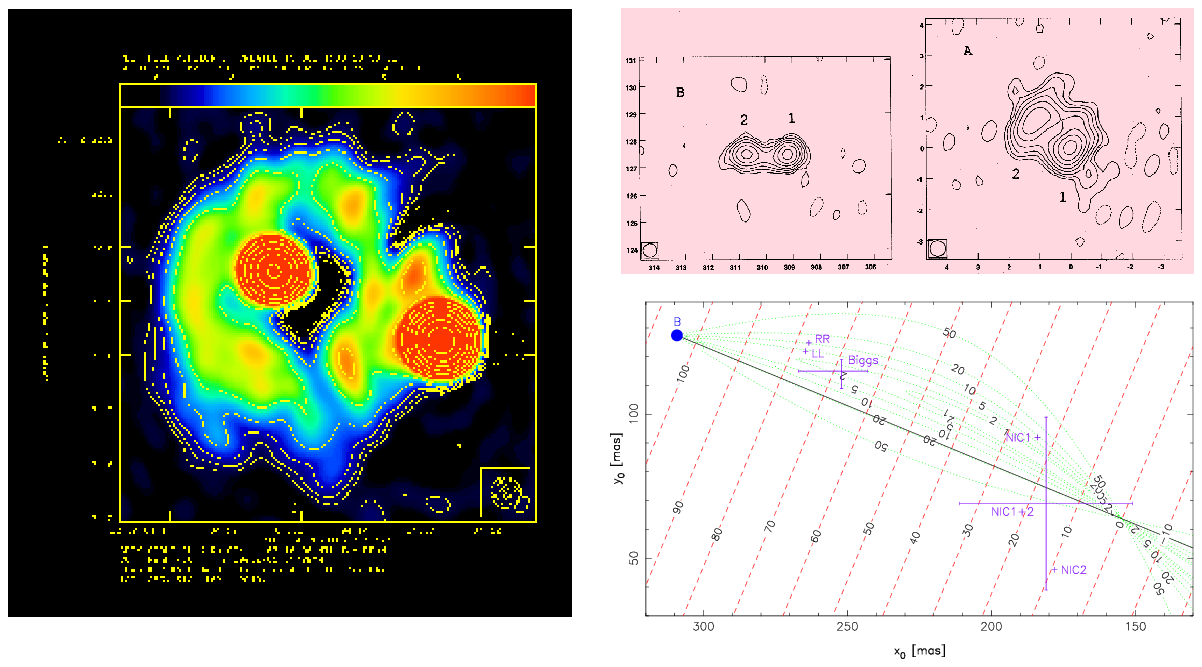}
\else%
\plotone{fig0218.eps}
\fi%
\fi%
\end{minipage}\end{center}
\begin{quote}Figure 1.\quad
a) 5\,GHz MERLIN/VLA map of the ring (Biggs et al.~2001)
  b) 15\,GHz VLBI maps of A and B (Patnaik, Porcas, and
  Browne 1995). c)  Contours of $H_0$ (straight lines) and $\chi^2$
residuals (dotted; without using the ring data) as a function of the
  lens position 
\end{quote}

Many constraints for the lens models are available from observations:
small scale substructure of the compact images on a 
scale of milliarcseconds (constrains the radial mass profile, which is
close to isothermal) and a highly structured Einstein ring. The
lens is an isolated non-interacting galaxy; lens models can thus be
kept simple (Leh{\'a}r et al.~2000).
The only disadvantage is the very small size of the system
($0\farcs33$). A direct measurement of the lens centre's position has
not been achieved yet, but will soon be possible (HST-ACS observations
in cycle~11, PI: Neal Jackson).

\section{LensCLEAN}

We used an improved version of the LensCLEAN algorithm (Kochanek \&
Na\-ra\-yan 1992; Ellithorpe, Kochanek, \& Hewitt 1996) to constrain the
position of the lens centre, which is the most important model parameter.
LensCLEAN works by deconvolving radio interferometer data under the
constraint of a given lens model. The residuals are
then used to find the best fitting lens model.

\begin{center}\begin{minipage}{0.82\textwidth}
\ifbw%
\plotone{figlc_bw.eps}
\else%
\plotone{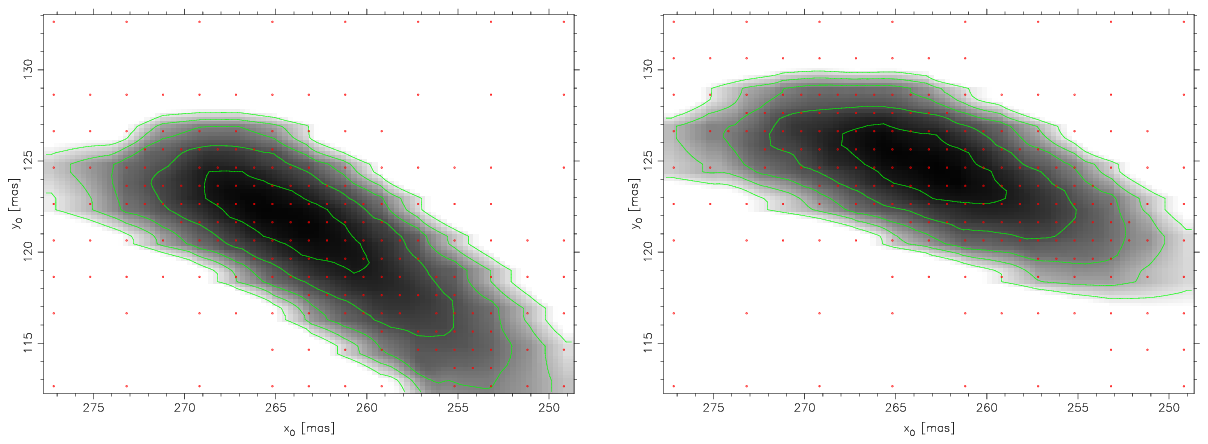}
\fi%
\end{minipage}\end{center}
\begin{quote}Figure 2.\quad
LensCLEAN residuals (VLA 15\,GHz) as a function of the lens position
(LL and RR polarization)
\end{quote}

The accuracy is of the
order a few mas, or a few per~cent in $H_0$. The result for singular
isothermal ellipsoidal models, using the time delay $\Delta t =
(10.5\pm0.4)$\,days from Biggs et al. (1999) is
\begin{displaymath}
H_0 = 77 \,{\rm km\,s^{-1}\,Mpc^{-1}} \quad \rm .
\end{displaymath}
Details will be published in an upcoming paper (Wucknitz et al., in prep.).
The original full-length poster version is available from the
author or on the web ({\tt http://www.hs.uni-hamburg.de/EN/Ins/Per/Wucknitz}).


\begin{references}
\reference Biggs, A.~D., Browne, I.~W.~A., Helbig, P., Koopmans,
  L.~V.~E., Wilkinson, P.~N., \& Perley, R.~A. 1999, \mnras, 304, 349
\reference Biggs, A.~D., Browne, I.~W.~A., Muxlow, T.~W.~B., \&
  Wilkinson, P.~N. 2001, \mnras, 322, 821
\reference Ellithorpe, J.~D., Kochanek, C.~S., \& Hewitt, J.~N. 1996,
  \apj, 464, 556
\reference Kochanek, C.~S., \& Narayan, R. 1992, \apj, 401, 461
\reference Leh{\'a}r, J., Falco, E.~E.,  Kochanek, C.~S., McLeod,
  B.~A., Mu{\~n}oz, J.~A., Impey, C.~D., Rix, H.~W., Keeton, C.~R., \&
  Peng, C.~Y. 2000, \apj, 536, 584
\reference Patnaik, A.~R., Porcas, R.~W., \& Browne, I.~W.~A. 1995,
  \mnras, 274, L5
\reference Refsdal, S. 1964, \mnras, 128, 307
\end{references}
\end{document}